\documentclass[%
floatfix,
reprint,
superscriptaddress,
nofootinbib,
amsmath,amssymb,
]{revtex4-1}
\pdfoutput=1
\usepackage{float}
\usepackage{xcolor}
\usepackage{graphicx}
\usepackage{dcolumn}
\usepackage{bm}
\usepackage{physics}
\usepackage[normalem]{ulem}
\usepackage[colorlinks=true, linkcolor=blue, citecolor=blue, urlcolor=blue]{hyperref}
\newcommand{\ZIB}{Zuse Institute Berlin, 14195 Berlin, Germany}
\newcommand{\JCM}{JCMwave GmbH, 14050 Berlin, Germany}

\usepackage{graphicx}
\usepackage{booktabs}
\usepackage{tabularx}

\begin{document}

\title{Resonance expansion of quadratic quantities with regularized quasinormal modes}

\author{Fridtjof Betz}
\affiliation{\ZIB}
\author{Felix Binkowski}
\affiliation{\ZIB}
\author{Martin Hammerschmidt}
\affiliation{\JCM}
\author{Lin Zschiedrich}
\affiliation{\JCM}
\author{Sven~Burger}
\affiliation{\ZIB}
\affiliation{\JCM}

\begin{abstract}
Resonance expansions are an intuitive approach to capture the interaction of an optical resonator with light. 
Here, we present a quasinormal mode expansion approach for quadratic observables exploiting the rigorous Riesz 
projection method. We demonstrate the approach by a numerical implementation of a state-of-the-art quantum 
light source and emphasize the ability of the approach to provide modal expansions outside the underlying nanophotonic resonator.
\end{abstract}

\maketitle

\section{Introduction}
\label{sec:introduction}
Resonances appear in all areas of physics where wave phenomena are involved, e.g., 
in nanophotonics~\cite{Lukyanchuk_Nat_Mat_2010,Novotny_NatPhot_2011},
in optomechanics~\cite{Aspelmeyer_RevModPhys_2014},
in quantum physics~\cite{Zworski_Resonances_1999,Tame_QuantPlasmon_NatPhys_2013}, 
or in acoustics~\cite{Cummer_Acoustics_2016}.
The theoretical description of the interaction of the resonances with driving source terms
is essential for design and fabrication of devices~\cite{West_LasPhotRev_2010,Lindquist_2012}.
Resonance expansion approaches can provide the necessary insights, 
since physical observables can often be decomposed into a few weighted resonance modes~\cite{Lalanne_QNMReview_2018,Nicolet_2022}.

In the field of nanophotonics, the resonances are given by electromagnetic fields,
the so-called eigenmodes, and corresponding eigenfrequencies.
The resonances are solutions to Maxwell's equations without a source term.
The resulting eigenproblems are
non-Hermitian because the systems exhibit losses due to open boundaries or due to damping losses~\cite{Lalanne_QNMReview_2018}.
This leads to complex-valued eigenfrequencies and spatially diverging eigenmodes.
In this context, the eigenmodes are also called quasinormal modes (QNMs). 
Various approaches for resonance expansion based on these QNMs have recently been 
intensively investigated~\cite{Muljarov_EPL_2010,Sauvan_QNMexpansionPurcell_2013,Zolla_OptLett_2018,Weiss_PRB_2018,Zschiedrich_PRA_2018,Franke_PRL_2019,Wu_2021_OSA}.

Optical observables are usually quadratic in the electric field.
An example is the electromagnetic field energy absorption.
It is not straightforward to calculate 
the resonance expansion of such quadratic quantities
in the near field of a corresponding nanostructure,
since cross terms can occur between the individual expansion terms~\cite{Bai_Optica_2013}.
Such terms would prevent the accurate examination of the influence of each single QNM in the expansion.
A further challenge is the resonance expansion of observables outside the nanoresonator, such as
the energy flux density in the far field. These quantities
can not only be quadratic, also the divergence of the QNMs with an increasing
distance to the resonator~\cite{Lalanne_QNMReview_2018, Kristensen_QNM_2020} has to be addressed properly.
Contour integration in the complex frequency plane allows to
calculate resonance expansions rigorously, for quadratic quantities near and far
from the nanoresonators and without the consideration of cross terms, which has been
shown recently~\cite{Binkowski_2020_PRB}. 
The Riesz projection (RP) approach, sketched in the left column of Fig.~\ref{fig:1},
accesses modal contributions of quadratic quantities by contour integrals
regarding the eigenfrequencies and the complex conjugated eigenfrequencies of the QNMs~\cite{Binkowski_2020_PRB}.

In this work, we propose a new approach for the QNM expansion of
quadratic quantities based on the same mathematical ideas. 
The approach, outlined in the right column of Fig.~\ref{fig:1}, does not exhibit cross terms and,
moreover, the field distributions of the QNMs can be used to expand quantities outside the nanoresonators
due to a regularization with solutions of scattering problems at the complex
conjugate eigenfrequencies of the QNMs. 
We demonstrate the applicability of this new approach by the modal analysis of various properties of a localized emitter in a circular Bragg grating (CBG). 
Resonance expansions of the field energy absorption in the near field
and of the energy flux in the near field and in the far field of the CBG are performed.

\section{Theoretical background}
\label{sec:Theory}

\begin{figure*}
\includegraphics{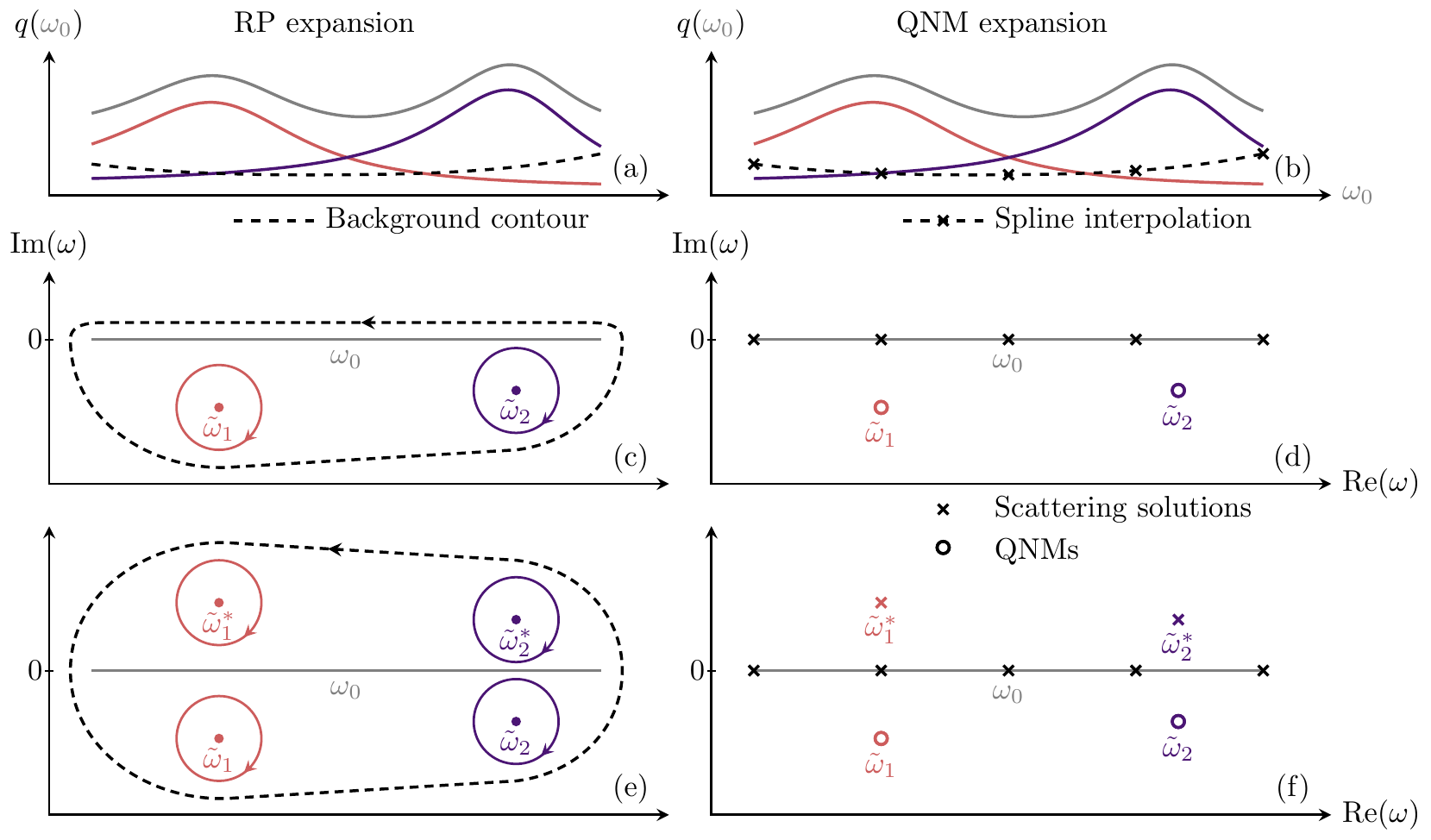}
\caption{Riesz projection (RP) expansion and quasi normal mode (QNM) expansion.
(a,b) The contributions of two resonances, associated with the eigenfrequencies $\tilde{\omega}_1$ and $\tilde{\omega}_2$, along with a background curve add up to the observable $q(\omega_0)$ as a function of real frequencies $\omega_0$.
(c) Integrals along the contours enclosing $\tilde{\omega}_1$ and $\tilde{\omega}_2$ provide the modal 
contributions to the RP expansion of observables $q(\omega_0)$ which linearly depend on the electric field. The background contribution is based on the background contour enclosing both eigenfrequencies.
(d) The respective QNM expansion requires coefficients $\alpha_n(\omega_0)$ to obtain $q(\omega_0)$ as a 
weighted sum of the QNMs. Cauchy's residue theorem relates the QNM expansion terms to the contour 
integrals of the RP expansion. The background contribution can be interpolated using solutions of the 
scattering problem at real frequencies.
(e) If $q(\omega_0)$ is quadratic in the electric field, additional contours around the complex conjugated eigenfrequencies $\tilde{\omega}_n^*$ are used in the RP formalism.
(f) For an exact QNM expansion of quantities $q(\omega_0)$ quadratic in the electric field, 
in addition to the QNMs solutions of scattering problems at the complex conjugated eigenfrequencies $\tilde{\omega}_n^*$ are required for the modal shares.} 
\label{fig:1} 
\end{figure*}

This section starts with an introduction of the relevant optical quantities. Basic concepts from QNM theory and the RP method will be recapitulated in Sec. \ref{sc:qnm} and Sec. \ref{sc:rp}, respectively. The new approach of QNM expansions of quadratic quantities is introduced in Sec.~\ref{sc:qf}. Based on the decomposition of a quantity $q(\omega_0)$ into two modal shares and a slowly varying background contribution, Fig.~\ref{fig:1} summarizes the ideas detailed below. Figure~\ref{fig:1}(a) and (b) outline the approach to RP expansion and QNM expansion, respectively, Fig.~\ref{fig:1}(c) and (d) show how to access the corresponding expansions of linear observables, and Fig.~\ref{fig:1}(e) and (f) illustrate the procedure for computing quadratic observables.

Hereafter, we assume a harmonic time dependence and linear materials. Consequently, without loss of generality, we can focus on observables $q(\omega_0)$ which can be written in terms of the electric field strength $\vb{E}(\vb{r},\omega_0) \in \mathbb{C}^3$. Besides the dependence on the angular frequency $\omega_0 \in \mathbb{R}$, there is a dependence on the position $\vb{r} \in \mathbb{R}^3$, which we drop for a shorter notation. Under these conditions, the second order Maxwell's wave equation 
\begin{equation} \label{eq:mxwl}
    \curl{\mu(\omega_0)^{-1}\curl{\vb{E}(\omega_0)}}-\omega_0^2\epsilon(\omega_0)\vb{E}(\omega_0) = i\omega_0\vb{J}
\end{equation}
describes a physical system subjected to a driving source $\vb{J} \in \mathbb{C}^3$. Material properties enter the equation with the permeability tensor $\mu(\omega_0)$, which for optical frequencies is typically equal to the vacuum permeability $\mu_0$, and the permittivity tensor $\epsilon(\omega_0)$. 

Solutions $\vb{E}(\omega_0)$ of Eq.~(\ref{eq:mxwl}) are referred to as solutions of scattering problems. The restriction in its formulation to real frequencies $\omega_0$ is equivalent to demanding physically relevant solutions. As we are interested in eigenmodes of open systems, i.e., systems which exchange energy with the environment, we will need to deal with solutions $\vb{E}(\omega)$ at complex frequencies $\omega \in \mathbb{C}$. These can be regarded as analytic continuations of $\vb{E}(\omega_0)$ into the complex plane. As a consequence, interpolated data from experiments does not suffice to represent material dispersion. 
Instead, fits to appropriate models which can be evaluated at complex frequencies, e.g., Drude-Lorentz models, are required. For this purpose specialized fitting routines are available~\cite{Sehmi_2017,Garcia-Vergara_2017}.

Measurable observables derived from the electric field usually depend quadratically on it and can be expressed as sesquilinear maps. We use the notation $\langle \vb{a},\vb{b}\rangle_q$ for a sesquilinear map corresponding to the quantity $q$. 
Additivity holds in both arguments and, furthermore, we have $\langle \alpha \, \vb{a} ,\beta \, \vb{b} \rangle_q= \alpha^\ast \, \beta \, \langle \vb{a}, \vb{b} \rangle_q$, where $\vb{a},\vb{b}\in\mathbb{C}^3$, $\alpha,\beta \in \mathbb{C}$ and the asterisk denotes the complex conjugate. Examples are the power $P_\mathrm{rad}$ dissipated into a solid angle $\Omega_\mathrm{NA}$ corresponding to a given numerical aperture (NA), 
\begin{equation} \label{eq:p_rad}
    \langle \vb{E},\vb{E}\rangle_{P_\mathrm{rad}} = \frac{1}{2} \int_{\Omega_\mathrm{NA}} \sqrt{\frac{\epsilon}{\mu_0}} \, \vb{E}^\ast \vdot \vb{E} \, \dd \Omega,
\end{equation} 
and the power $P_\mathrm{abs}$ dissipated within a volume $V$ due to absorption,
\begin{equation} \label{eq:p_abs}
    \langle\vb{E},\vb{E}\rangle_{P_\mathrm{abs}} = -\frac{\omega_0}{2}\Im\left[\int_V (\vb{D}^\ast\vdot\vb{E}+\vb{H}^\ast\vdot\vb{B})\,\dd V\right].
\end{equation}
In the last example, we have introduced the electric flux density $\vb{D} = \epsilon\vb{E}$, the magnetic flux density $\vb{B} = 1/i\omega_0\curl{\vb{E}}$, and the magnetic field strength $\vb{H} = 1/\mu_0\vb{B}$.

\subsection{QNM expansion} \label{sc:qnm}
This section summarizes how the ingredients indicated in Fig.~\ref{fig:1}(d) give rise to the QNM expansion in Fig.~\ref{fig:1}(b) if $q(\omega_0)$ is linear in $\vb{E}(\omega_0)$ \cite{Wu_2021_OSA,Sauvan_QNMexpansionPurcell_2013}.

Setting the source $\vb{J}$ in Eq.~(\ref{eq:mxwl}) to zero yields the eigenproblem
\begin{equation} \label{eq:evs}
    \curl{\mu(\tilde{\omega}_n)^{-1}\curl{\vb{\tilde{E}}_n(\tilde{\omega}_n)}}-\tilde{\omega}_n^2\epsilon(\tilde{\omega}_n)\vb{\tilde{E}}_n(\tilde{\omega}_n) = 0.
\end{equation}
If absorptive materials or transparent boundary conditions introduce dissipation into Eq.~(\ref{eq:evs}), the eigenfrequencies $\tilde{\omega}_n \in \mathbb{C}$ have negative imaginary parts which account for an exponential decay in time and cause the corresponding eigenmodes to diverge in space. For this reason, normalization requires special care and the normalized resonances are referred to as QNMs.

Numerical results indicate that QNMs form a complete basis of the electric field inside an arbitrary resonator placed in a homogeneous background material~\cite{Lalanne_QNMReview_2018,Kristensen_QNM_2020}. These findings were preceded by theoretical results which proved the completeness for simple symmetric geometries~\cite{Leung_CompleteOrthQNM_1994}. Hence, solutions of the scattering problem can be written as the weighted sum (QNM expansion),
\begin{equation} \label{eq:qnm_expansion}
    \vb{E}(\omega_0) = \vb{R}(\omega_0) + \sum_{n=1}^{N} \alpha_n(\omega_0) \vb{\tilde{E}}_n,
\end{equation}
of $N$ QNMs $\vb{\tilde{E}}_n$ with expansion coefficients $\alpha_n(\omega_0)$ plus a residual contribution $\vb{R}(\omega_0)$ that vanishes for large $N$. If the background is not homogeneous and the QNM basis is therefore not complete, $\vb{R}(\omega_0)$ can be extended to include non-resonant contributions. As sketched in the right column of Fig.~\ref{fig:1}, $\vb{R}(\omega_0)$ can be evaluated using solutions of scattering problems at real frequencies \cite{Wu_2021_OSA}, e.g., by interpolating $\vb{R}(\omega_0) = \vb{E}(\omega_0)-\sum_{n=1}^N \alpha_n(\omega_0)\vb{\tilde{E}}_n$ with splines.

The expansion coefficients are available as closed form expressions for arbitrary sources~\cite{Lalanne_QNMReview_2018} and appropriate near-field to far-field transformations \cite{Yang_ACS_Phot_2016,Ren_PRB_2020} provide approximate values for the field far away from the resonator. For our purpose, it suffices to write down the expansion coefficients for a point dipole source at position $\vb{r}_0$, with dipole moment $\vb{p}$ and strength vector $\vb{j} = -i\omega_0\vb{p}$. Modeling it using an electric current density
\begin{equation} \label{eq:dipole}
    \vb{J}(\omega_0) = \vb{j}\delta(\vb{r}-\vb{r}_0),
\end{equation}
where $\delta$ is the Dirac delta distribution, we get \cite{Sauvan_QNMexpansionPurcell_2013}
\begin{equation}
    \alpha_n(\omega_0) = \frac{i}{\tilde{\omega}_n-\omega_0}\vb{\tilde{E}}_n(\vb{r}_0)\vdot \vb{j}.
    \label{eq:coeff}
\end{equation} 

Equation~(\ref{eq:qnm_expansion}) allows to expand any quantity derived from solutions to the scattering problem. Yet, in general, if they appear squared, all the cross terms have to be considered and the contributions of individual modes cannot be discriminated. However, for some quadratic quantities, alternative expressions exist which are linear in $\vb{E}(\omega_0)$. A prominent example is the Purcell enhancement $\Gamma(\omega_0)$ acting on the dipole emitter defined in Eq.~(\ref{eq:dipole}). $\Gamma(\omega_0)$ is the quotient of the total power emitted by the dipole and the power emitted in homogeneous bulk material $\Gamma_\mathrm{b}$. One can write
\begin{align}
    \Gamma(\omega_0) =& \frac{1}{2}\Re\qty[\int_{S}\vb{E}(\omega_0)\cross\vb{H}^\ast(\omega_0)\vdot \vb{n} \dd{S}]/\Gamma_\mathrm{b} 
    \label{eq:ndk_flux}\\
    =& -\frac{1}{2}\Re\qty[\vb{E}(\vb{r}_0,\omega_0)\vdot\vb{j}^\ast]/\Gamma_\mathrm{b} 
    \label{eq:ndk_point}
\end{align}

if the surface $S$ with unit normal $\vb{n}$ in Eq.~(\ref{eq:ndk_flux}) encloses a non-absorbing surroundings of the source. Equation~(\ref{eq:ndk_point}) follows after using Eq.~(\ref{eq:mxwl}) and integration by parts to transform the flux integral of the complex Poynting vector, $\vb{E}(\omega_0)\cross\vb{H}^\ast(\omega_0) = \vb{E}(\omega_0)\cross\frac{1}{\mu_0}\curl{\vb{E}^\ast(\omega_0)}$, into a volume integral. Furthermore, the source term given in Eq.~(\ref{eq:dipole}) reduces the volume integral to evaluating $\vb{E}(\vb{r}_0,\omega_0)$ at the position $\vb{r}_0$ of the dipole. Substituting Eq.~(\ref{eq:qnm_expansion}) into Eq.~(\ref{eq:ndk_point}) yields the modal expansion of $\Gamma(\omega_0)$.

\subsection{Riesz projection expansion} \label{sc:rp}

In general, the RP expansion of the electric field is given by
\begin{align}
\begin{split}
    \vb{E}(\omega_0) = \frac{1}{2 \pi i}&\oint\limits_{C_\mathrm{bg}}\frac{\vb{E}(\omega)}{\omega-\omega_0}\dd\omega \\
    &-\sum_{n=1}^N\frac{1}{2\pi i}\oint\limits_{\tilde{C}_n}\frac{\vb{E}(\omega)}{\omega-\omega_0}\dd\omega.
\end{split}\label{eq:rp_expansion}
\end{align}
Here, instead of QNMs with associated expansion coefficients, contour integrals provide the contributions of individual eigenmodes. Hence, neither normalization nor completeness is required. The integral along the background contour $C_\mathrm{bg}$ gives an explicit expression for the residual term $\vb{R}(\omega_0)$ that includes possible non-resonant contributions.
The schematic in Figure~\ref{fig:1}(c) shows corresponding contours in the complex frequency plane, 
representing a RP with $N=2$. 

For the expansion of sesquilinear maps $\langle\vb{E}(\omega_0),\vb{E}(\omega_0)\rangle_q$ both $\vb{E}^\ast(\omega_0)$ and $\vb{E}(\omega_0)$, are analytically continued to the complex plane \cite{Binkowski_2020_PRB}. Hence, it becomes necessary to introduce additional contours enclosing the complex conjugate eigenfrequencies $\omega_n^\ast$ of the QNMs as sketched in Fig.~\ref{fig:1}(e). 
Any specific orthogonality relation between the eigenmodes is not required, and nevertheless, the resulting expansion does not contain cross terms. 
The contribution corresponding to a given eigenfrequency $\omega_m$
\begin{align} \label{eq:qf_contribution}
\begin{split}
    q_m(\omega_0) = -\frac{1}{2\pi i}&\oint\limits_{\tilde{C}_m} \frac{\langle\vb{E}(\omega^\ast),\vb{E}(\omega)\rangle_q}{\omega-\omega_0}\, \dd\omega \\&- \frac{1}{2\pi i}\oint\limits_{\tilde{C}_m^\ast} \frac{\langle\vb{E}(\omega^\ast),\vb{E}(\omega)\rangle_q}{\omega-\omega_0}\, \dd\omega,
\end{split}
\end{align}
is the sum of two integrals along $\tilde{C}_m$ and $\tilde{C}_m^\ast$ enclosing $\tilde{\omega}_m$ and its complex conjugate $\tilde{\omega}_m^\ast$, respectively.

\begin{figure*}
\includegraphics{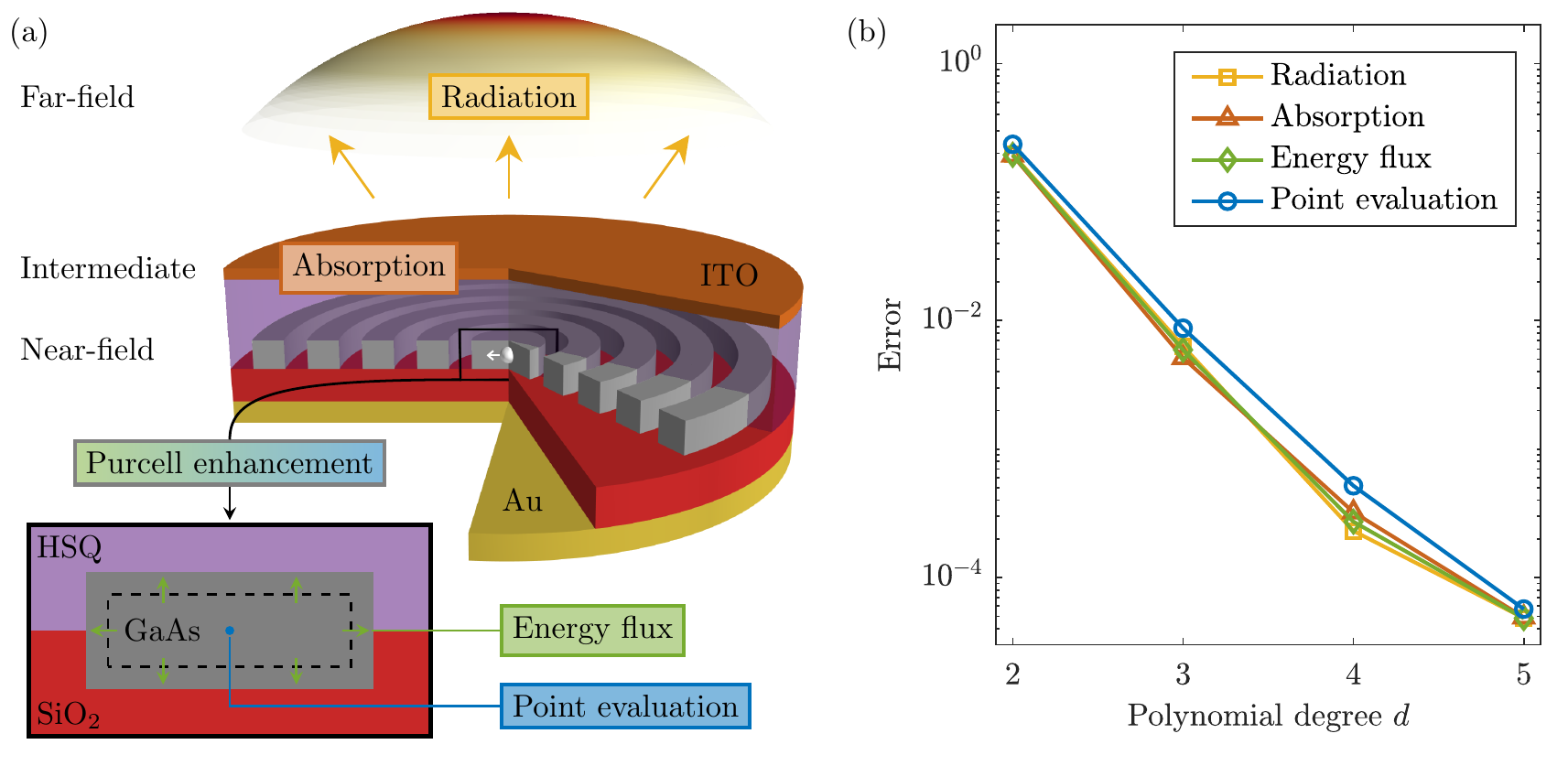}
\caption{Computing modal contributions to selected quantities. (a) Schematic of a circular Bragg grating above a gold mirror, with a point dipole source in the central disk. The regions where the quantities of interest are evaluated are pointed out. The radiated power is integrated over a numerical aperture of $\mathrm{NA} = 0.6$ and belongs to the far field. The absorption is calculated within the ITO layer. The magnification of the central disk sketches two methods commonly used to evaluate the Purcell enhancement: the energy flux through a surface enclosing the emitter and the point evaluation of the electric field at its position. (b) Convergence of the dominant eigenmode contribution to the various quantities. 
The error is defined as $\varepsilon = \max(q_d-q_\mathrm{ref})/q_\mathrm{max}$. Here, the contributions $q_d$ of the quantity $q$ are evaluated using quasi normal mode expansion of quadratic forms as indicated in Fig.~\ref{fig:1}(f). The numerical accuracy of $q_d$ is improved by increasing the polynomial degree $d$ in the finite-element discretization of the numerical approximation. The reference solution $q_\mathrm{ref}$ uses the Riesz projection method outlined in Fig.~\ref{fig:1}(e), and is of polynomial degree $d=7$. The normalization $q_\mathrm{max}$ is the maximal value of the reference solution.}
\label{fig:2} 
\end{figure*}

\subsection{QNM expansion of quadratic forms} \label{sc:qf}

The main subject of this manuscript is to provide an expression for modal contributions to quadratic quantities in terms of QNMs. This will be achieved evaluating the contour integrals in Eq.~(\ref{eq:qf_contribution}) analytically. 

To facilitate the notation we start with the corresponding linear expression. We extend the expansion of $\vb{E}(\omega_0)$, Eq.~(\ref{eq:qnm_expansion}), to the complex plane imposing no restriction on $\vb{R}(\omega)$ other than that it must be analytic within the compact domain defined by the background contour in Fig.~\ref{fig:1}(e). With the resulting expansion of $\vb{E}(\omega)$ the contribution to the RP expansion associated to the eigenfrequency $\tilde{\omega}_m$ is
\begin{align*}
    -\frac{1}{2\pi i}\oint\limits_{\tilde{C}_m}\frac{\vb{E(\omega)}}{\omega-\omega_0}\,\dd\omega =& -\frac{1}{2\pi i} \oint\limits_{\tilde{C}_m^1}\frac{\vb{R}(\omega)}{(\omega-\omega_0)}\,\dd\omega\nonumber\\
    &- \sum_{n=1}^{N} \frac{1}{2\pi i}\oint\limits_{\tilde{C}_m}\frac{\alpha_n(\omega) \vb{\tilde{E}}_n}{(\omega-\omega_0)}\,\dd\omega
    \, .
\end{align*}
All summands with $n\neq m$ and the integral with $\vb{R}(\omega)$ are analytic inside $\tilde{C}_m$, 
that can be chosen arbitrarily close to $\tilde{\omega}_n$ and can be dropped. With Cauchy's residue theorem we get from the remaining term $-\frac{1}{2\pi i}\oint_{\tilde{C}_m}\frac{\alpha_m(\omega) \vb{\tilde{E}}_m}{(\omega-\omega_0)}\,\dd\omega$
\begin{equation}
    -\frac{1}{2\pi i}\oint\limits_{\tilde{C}_m}\frac{\vb{E(\omega)}}{\omega-\omega_0}\,\dd\omega =
     \,\alpha_m(\omega_0)\vb{\tilde{E}}_m \, . \label{eq:rp_qnm}
\end{equation}

Similarly, we can introduce QNMs into Eq.~(\ref{eq:qf_contribution}). We note that $\vb{E}(\omega^\ast)$ is analytic inside $\tilde{C}_m$ but has a pole at $\omega = \tilde{\omega}_m^\ast$ inside $\tilde{C}_m^\ast$. The contour integrals in the expression for $q_n(\omega_0)$ noted in Eq.~(\ref{eq:qf_contribution}) can be evaluated based on Eq.~(\ref{eq:rp_qnm}) and we get the expression for the modal shares,
\begin{align}
\begin{split}
    q_m(\omega_0) = \alpha_m(\omega_0)&\langle\vb{E}(\omega_m^\ast),\vb{\tilde{E}}_m\rangle_q \\
    &+ \alpha_m^\ast(\omega_0)\langle\vb{\tilde{E}}_m,\vb{E}(\omega_m^\ast)\rangle_q\, ,
\end{split} \label{eq:qf_contribution_qnm}
\end{align}
 which depends on QNMs and is regularized by scattering problems at the complex conjugated eigenfrequencies $\omega_n^\ast$. For this reason, the expression is valid at any distance from the resonator. 
Fig.~\ref{fig:1}(f) shows schematically how the method of exact QNM expansion relies on appropriately normalized QNMs
as well as on solutions of scattering problems at $\tilde{\omega}_n^*$.
 With this result we can eventually write down the expansion for a quadratic quantity in terms of QNMs,
 \begin{equation}
     q(\omega_0) = q_\mathrm{r} + \sum_{n=1}^N q_n(\omega_0).
     \label{eq:qnm_expansion_quadratic_forms}
 \end{equation}
 The residual contribution $q_\mathrm{r}$ can either be evaluated using scattering simulations at real frequencies (Sec.~\ref{sc:qnm}) or a background contour (Sec.~\ref{sc:rp}).

\section{Application}

We perform numerical experiments on the example of a hybrid circular Bragg grating (hCBG) which acts as a nanoantenna 
for an efficient single-photon source, based on a recently proposed design~\cite{Rickert_FC-elCBG_2022}.
The circular Bragg grating is patterned into a gallium arsenide (GaAs) 
slab and allows to employ, e.g., self-assembled quantum dots as emitters.
It is placed on a silicon dioxide (SiO\textsubscript{2}) spacer inside a capacitor, composed of a gold mirror and a transparent indium tin oxide (ITO) top contact. 
The space between hCBG and ITO and between the grating rings is filled with hydrogen silsesquioxane (HSQ).
Figure~\ref{fig:2}(a) provides a schematic of the physical system, that is driven by an $x$-polarized dipole source placed in the center of the inner disk, modeled with Eq.~(\ref{eq:dipole}) with $|\vb{j}| = 0.2\,\mu \textrm{A}/\textrm{m}^2$.
The optical properties of the various materials are obtained from tabulated data (see Ref. \cite{Rickert_FC-elCBG_2022} Tab. 3).

The numerical realization relies on the open source software   \mbox{\textsc{RPExpand}}~\cite{Betz_2021}, 
which is a \textsc{MATLAB} package for modal expansions of physical observables in resonant systems. 
\mbox{\textsc{RPExpand}} has been extended to support QNM expansions. 
A recent version of \textsc{RPExpand} together with geometry and material parameters is provided in the 
related data publication~\cite{Betz_SourceCode_zenodo}.
The numerical discretization of Maxwell's equations relies on the commercial solver 
\textsc{JCMsuite} which implements the finite element method (FEM) for solving 
scattering and eigenproblems~\cite{Pomplum_NanoopticFEM_2007} with perfectly matched 
layers (PML)~\cite{Berenger_1994} to account for open boundary conditions and to 
normalize the QNMs. 
The computational effort is reduced by exploiting the rotational symmetry of the system.

\begin{figure}
\includegraphics{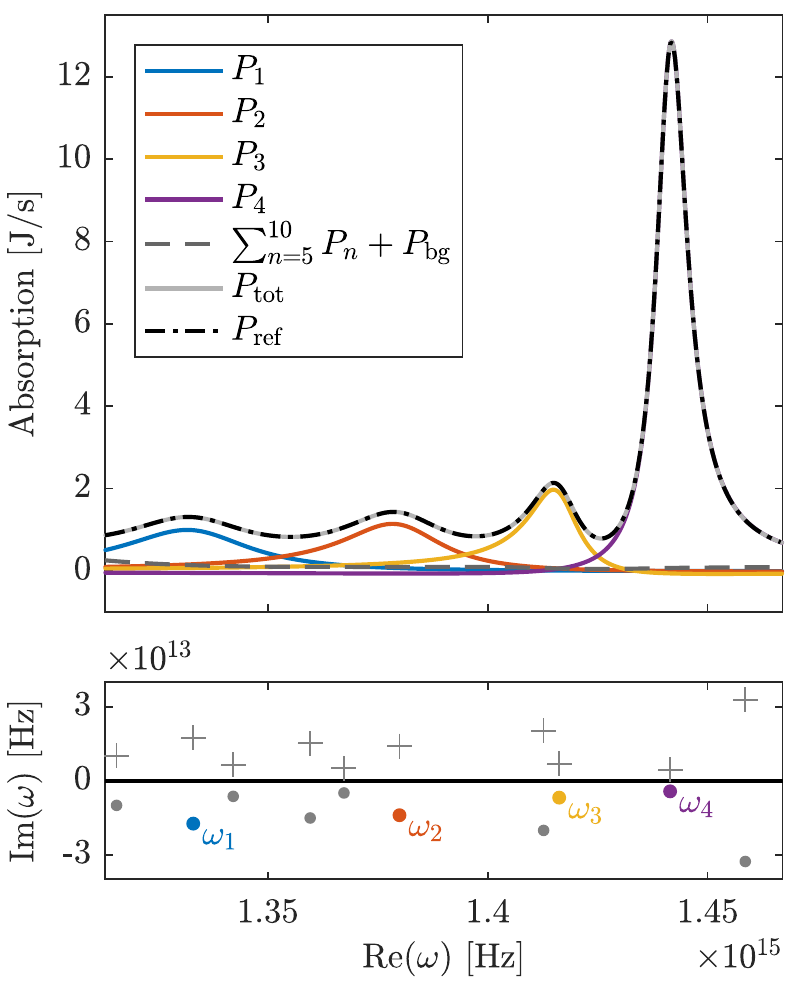}
\caption{Modal expansion of the electromagnetic field absorption in the ITO layer located above the circular Bragg grating schematically shown in Fig.~\ref{fig:2}(a). Four modes, whose positions in the complex plane are marked in the lower part of the figure, dominate the spectrum. The background contribution $P_\mathrm{bg}$ relies on cubic spline interpolation connecting 20 values from real frequency simulations, equally spaced within the interval of interest.}
\label{fig:3} 
\end{figure}

First, the quasinormal mode expansion of quadratic forms is validated by investigating the numerical convergence of 
various quadratic quantities towards corresponding quasi-exact results. 
The modal contribution of the dominant mode ($\omega_4$ in Fig.~\ref{fig:3}) to the following quantities is investigated:
the power radiated into a numerical aperture of NA=0.6 in the far field, Eq.~(\ref{eq:p_rad}), 
the absorption in the ITO layer, Eq.~(\ref{eq:p_abs}), 
and the Purcell enhancement based on a flux integral, Eq.~(\ref{eq:ndk_flux}), and based on point
evaluation, Eq.~(\ref{eq:ndk_point}).
The numerical accuracy is varied by increasing the polynomial degree $d$ of the FEM discretizations of the model, from $d=2$ to $d=5$. 
The quasi-exact results, $q_\textrm{ref}$, are obtained using the RP expansion of quadratic forms, Eq.~(\ref{eq:qf_contribution}), with a FEM discretization with $d=7$.
As shown in Fig.~\ref{fig:2}(b), the results of the QNM expansion converge to the results of the RP expansion, 
with very good achieved numerical accuracies, which compare to the numerical accuracies of the involved scattering and 
eigenmode problems. 

Next, QNM expansion of field absorption in a region outside of the resonator is demonstrated. 
The investigated setup supports 10 eigenmodes in the chosen subset of the complex frequency plane, depicted in the lower part of Fig.~\ref{fig:3}. 
As shown in the upper part of Fig.~\ref{fig:3}, four of these modes significantly couple to the given source and shape the absorption spectrum. 
The remaining modal shares are close to zero and form together with the background contribution $P_\mathrm{bg}$ 
a slowly varying offset. 
For the background contribution $P_\mathrm{bg}$, the nodes $P_{\mathrm{bg,m}} = P_\mathrm{ref}(\omega_{0,m})-\sum_{n=1}^{10}P_n(\omega_{0,m})$ have been interpolated with cubic splines. 
The index $m$ runs from 1 to 20 and the real frequencies $\omega_{0,m}$ are equidistantly distributed over the interval of interest. 
A comparison to a frequency scan with 308 scattering solutions results in a maximum relative error of approximately $5\times10^{-4}$. 
This result is consistent with the convergence analysis in Fig.~\ref{fig:2}(b).

\section{Discussion}
After validating the QNM expansion of quadratic forms and demonstrating its applicability, general observations and remarks are
given. 

\begin{figure}
\includegraphics{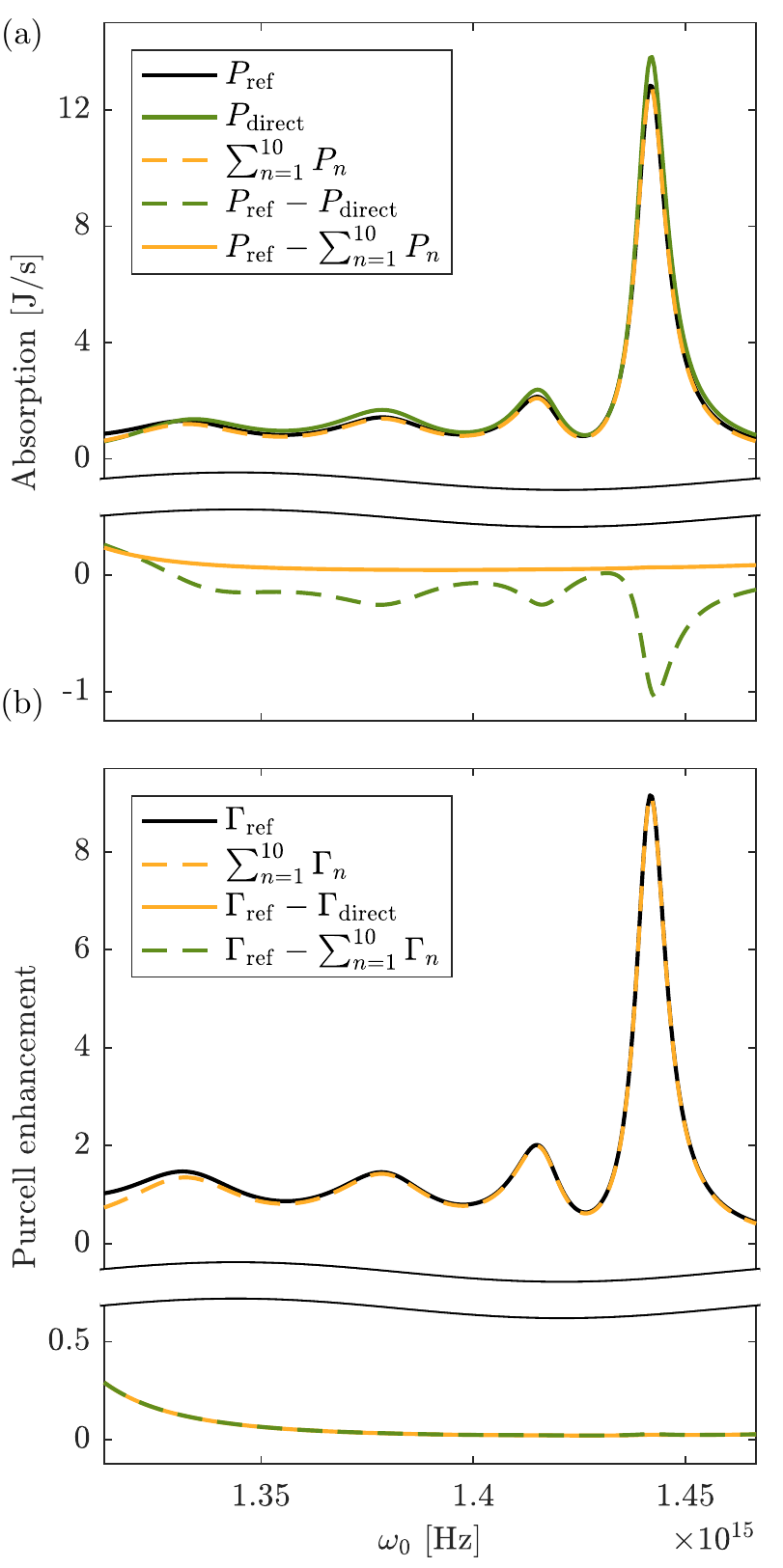}
\caption{Comparison between modal expansions based on the proposed method, $\sum_{n=1}^{10}P_n$, and $P_{\mathrm{direct}}= \frac{1}{2}\sum_{n,m=1}^{10}\vb{\tilde{E}}_n \vdot \vb{\tilde{D}}_m$, respectively. (a) The negative imaginary parts of the eigenfrequencies lead to an overestimation of the absorption in $P_\mathrm{direct}$ as the absorbing layer is located outside the resonator. The contributions $P_n$, defined in Eq.~(\ref{eq:qf_contribution_qnm}), do not suffer from the exponential growth. The lower part shows magnified differences between reference solutions $P_\mathrm{ref}$ obtained from real frequency simulations and the modal expansions $P_{\mathrm{direct}}$ and $\sum_{n=1}^{10}P_n$. (b) The Purcell enhancement is evaluated using a point evaluation at the dipole position ($\Gamma_{\mathrm{direct}}$) or an integral of the energy flux over a closed surface enclosing the emitter ($\Gamma_\mathrm{ref}$ and $\Gamma_n$). Here, the fields are evaluated in the center of the resonator. As expected, both methods coincide and the smooth background can be easily interpolated.}
\label{fig:4} 
\end{figure}

Firstly, the QNM expansion method inherits from the RP method its ability to provide modal expansions of physical observables at any distance from the resonator. 
Far-field quantities have been discussed in a previous work~\cite{Binkowski_2020_PRB}. 
Here, we point out that numerical experiments, presented in Fig.~\ref{fig:4}(a), 
indicate that the modal contributions to the absorption in the ITO layer are overestimated if the QNMs 
are not regularized, i.e., if the sum from Eq.~(\ref{eq:qnm_expansion}) is directly inserted into Eq.~(\ref{eq:p_abs}). 
The consequence is a background contribution that exhibits negative peaks at resonant frequencies. These features render accurate interpolation with few control points infeasible. 
These problems do not arise when the absorption is expanded in a region where the divergence of the modal fields is not relevant, i.e., inside the resonator.
For example, the absorption cross section of a gold nano rod has been successfully expanded using QNMs~\cite{Bai_Optica_2013}. In the given example, in contrast, the presence of the ITO coating does hardly effect the dominant resonances of the system and must be considered outside the resonator. In addition, it should be noted that the QNMs might not form a complete basis~\cite{Lalanne_QNMReview_2018} due to the layered background. However, as long as the background contribution is sufficiently flat, few scattering problems suffice for accurate interpolation.

The results for the Purcell enhancement $\Gamma(\omega_0)$ presented in Fig.~\ref{fig:4}(b) confirm the equivalence of the direct expansion in QNMs $\Gamma_\mathrm{QNM}$ and the expansion $\sum_{n=1}^{10}\Gamma_n(\omega_0)$ based on 
Eq.~\eqref{eq:qnm_expansion_quadratic_forms}
if the field is evaluated inside the resonator. 
Simultaneously, the expansion of $\Gamma(\omega_0)$ based on Eq.~(\ref{eq:ndk_flux}), which is quadratic in $\vb{E}(\omega_0)$ is compared to the expansion that results from Eq.~(\ref{eq:ndk_point}). For both expressions the field is evaluated inside the central disk of the resonator.

Further, it is worth elaborating on the relation between RP and QNM expansions expressed in Eq.~(\ref{eq:rp_qnm}). For the coefficients specified in Eq.~(\ref{eq:coeff}), the sums in Eq.~(\ref{eq:rp_expansion}) and Eq.~(\ref{eq:qnm_expansion}) are composed of the same modal shares. Yet, the expansion in QNMs is not unique~\cite{Gras_OSA_2020}. The same holds for RP expansions. Multiplying a function $c(\omega_0,\omega)$ to the integrands of the contour integrals yields different decompositions of $\vb{E}(\omega_0)$ if $c(\omega_0,\omega)$ is analytic in the compact region defined by the background contour and $c(\omega_0,\omega_0)=1$. The natural choice $c(\omega_0,\omega) = 1$ coincides with the expansion coefficients given in Eq.~(\ref{eq:coeff}). Another choice that gives rise to the expansion proposed in~\cite{Zolla_OptLett_2018} is $c_0 = \omega_0/\omega$. Similarly to Eq.~(\ref{eq:rp_qnm}) we get $-\frac{1}{2\pi i}\int_{C_m}\frac{\omega_0}{\omega}\frac{\vb{E}(\omega)}{\omega-\omega_0}\dd\omega = \alpha_m(\omega_0) \vb{\tilde{E}}(\omega)$ with $\alpha_m(\omega_0)= \frac{i\omega_0}{\omega_m(\tilde{\omega}_m-\omega_0)}\vb{\tilde{E}}_m(\vb{r}_0)\vdot \vb{j}$.

Eventually, we note that for computing QNMs, i.e., solutions to general nonlinear eigenvalue problems, 
both, 
direct methods like the Arnoldi and the Jacobi-Davidson method \cite{Voss_nonlinArnoldi_2004,Voss_nonlinJacobiDavid_2007}, 
and contour integration methods like the FEAST eigensolver~\cite{Gavin_JCompPhy_2018} 
or a RP based method~\cite{Binkowski_JCOMP_2020} can be used. 
The latter allows also for the direct computation of 
partial derivatives of eigenvalues~\cite{Binkowski2022_commun_phys}, which 
in principle allows for modal expansions of sensitivities in parameterized problem settings. 

\vspace{-0.19cm}
\section{Conclusions}
We have presented a formalism for expanding quadratic forms efficiently using few dominant resonances. The approach is based on the theory of the RP method and on using the field distributions of QNMs. It does not require contour integration and can be easily integrated in existing QNM frameworks such as the 
{\sc RPExpand}~\cite{Betz_2021} or the MAN~\cite{WU_MAN_Cumput_Phys_2022} programmes. 
For accurate results, the background contributions can be accounted for by interpolation relying either on a contour integral or scattering simulations at real frequencies. Furthermore, we have presented results showing that the exclusive use of QNMs for expansions outside the resonators can result in an overestimation of modal contributions and confirmed the equivalence of QNM and RP expansions in the given framework. 

\section*{Data and code availability}
Tabulated data and scripts for reproducing the presented numerical results are provided in the associated data publication, 
together with a current version of \mbox{\textsc{RPExpand}} that supports QNM expansion~\cite{Betz_SourceCode_zenodo}.

\section*{Acknowledgments}
We acknowledge funding by the German Federal Ministry of Education and Research (BMBF Forschungscampus MODAL, project 05M20ZBM) 
and by the Deutsche Forschungsgemeinschaft (DFG, German Research Foundation) under Germany’s Excellence Strategy - 
The Berlin Mathematics Research Center MATH+ (EXC-2046/1, project ID: 390685689). 
This project has received funding from the EMPIR programme co-financed by the Participating States and from the European Union’s Horizon 2020 research and innovation programme (project 20FUN02 POLIGHT). 
We thank Lucas Rickert and Tobias Heindel for discussions regarding the hCBG design investigated in this work.

%

\end{document}